\documentclass[showpacs,twocolumn,aps,preprintnumbers,letterpaper]{revtex4}
\usepackage{amsmath,amssymb}
\usepackage{epsfig}
\usepackage{graphicx}
\usepackage{amsmath}
\usepackage{slashed}
\usepackage{amsfonts}
\usepackage{epstopdf}

\newcommand{\be}{\begin{equation}}
\newcommand{\ee}{\end{equation}}
\newcommand{\bear}{\begin{eqnarray}}
\newcommand{\eear}{\end{eqnarray}}
\newcommand{\ba}{\begin{array}}
\newcommand{\ea}{\end{array}}

\def\be{\begin{eqnarray}}
\def\ee{\end{eqnarray}}
\def\bea{\be}
\def\eea{\ee}

\def\roughly#1{\mathrel{\raise.3ex\hbox{$#1$\kern-.75em%
\lower1ex\hbox{$\sim$}}}}

\begin{document}

\title{Heavy Exotic Molecules with  Charm and Bottom }

\author{Yizhuang Liu}
\email{yizhuang.liu@stonybrook.edu}
\affiliation{Department of Physics and Astronomy, Stony Brook University, Stony Brook, New York 11794-3800, USA}

\author{Ismail Zahed}
\email{ismail.zahed@stonybrook.edu}
\affiliation{Department of Physics and Astronomy, Stony Brook University, Stony Brook, New York 11794-3800, USA}


\date{\today}
\begin{abstract}
We revisit the formation of pion-mediated heavy-light  exotic molecules with both charm and bottom and their 
chiral partners under the general strictures of both heavy-quark and chiral symmetry. The chiral exotic partners with good
parity  formed using the $(0^+, 1^+)$ multiplet are about twice more bound than their primary exotic partners formed using
the $(0^-,1^-)$ multiplet.  The  chiral couplings across the multiplets $(0^\pm, 1^\pm)$  cause the chiral exotic
partners to unbind, and the primary exotic molecules to be about twice more bound, for   $J\leq 1$. Our multi-channel
coupling results show that only the charm isosinglet exotic molecules with  $J^{PC}=1^{++}$ binds,
which we identify as the reported neutral $X(3872)$. Also,  the bottom isotriplet exotic with $J^{PC}=1^{+-}$ binds, which we identify
as a mixture of the reported charged exotics $Z^+_b(10610)$ and $Z^+_b(10650)$. 
The bound isosinglet with $J^{PC}=1^{++}$ is suggested
as a possible neutral $X_b(10532)$ not yet reported.
\end{abstract}


\maketitle

\setcounter{footnote}{0}


\section{Introduction}
A decade ago, both the BaBar collaboration~\cite{BABAR}  and the CLEOII collaboration~\cite{CLEOII} 
have reported narrow peaks in the $D_s^+\pi^0$ (2317 MeV) and the $D_s^{*+}\pi^0$ (2460 GeV)
channels as expected from general chiral symmetry arguments~\cite{MACIEK,BARDEEN}. 
In QCD the light quark sector (u, d, s) is dominated by the spontaneous breaking of chiral
symmetry, while the heavy quark sector (c, b, t) is characterized by heavy-quark symmetry~\cite{ISGUR}. 
The combination of both symmetries led to the conclusion that the heavy-light
doublet $(0^-, 1^-)=(D,D^*)$ has a chiral partner $(0^+, 1^+)=(\tilde D,{\tilde D}^*)$ that is
about  one consituent mass heavier~\cite{MACIEK,BARDEEN}.

Recently, the Belle collaboration~\cite{BELLE} and the BESIII collaboration~\cite{BESIII} have reported the observations
of multiquark exotics. A major provider for these exotics is $\Upsilon(10860)$  and its ideal location near  the thresholds for 
$B\bar B^*\pi$ (10744) and $B^*\bar B^*\pi$ (10790) decays. The smallness of the  available phase space in the hadronic
decay of $\Upsilon(10860)$ calls for a compound with a long life-time, perhaps in a molecular configuration with heavy meson constituents. Several heavy exotic molecules
with quantum numbers uncommensurate with the excited states of charmonia and bottomia have been reported, 
such as the neutral $X(3872)$ and the charged $Z_c(3900)^\pm$ and $Z_b(10610)^\pm$. More of these exotics are expected to be unravelled by the DO collaboration at Fermilab~\cite{DO},  and the  LHCb collaboration at Cern~\cite{LHCb}.

Theoretical arguments have predicted the 
occurence of some of these exotics as molecular bound states mediated by one-pion exchange much
like deuterons or deusons~\cite{MOLECULES,THORSSON}. 
A number of molecular estimates regarding the occurence of doubly heavy exotic mesons with 
both charm and bottom content were suggested by many~\cite{THORSSON,KARLINER,OTHERS,OTHERSX,OTHERSZ,OTHERSXX}.
Non-molecular heavy exotics were also discussed using constituent quark models~\cite{MANOHAR}, 
heavy solitonic baryons~\cite{RISKA,MACIEK2}, instantons~\cite{MACIEK3} and QCD sum rules~\cite{SUHONG}. The molecular
mechanism favors the formation of shallow bound states near treshold, while the non-molecular
mechanism suggests the exitence of deeply bound states. The currently reported exotics by
the various experimental collaborations are in support of the molecular configurations.

The purpose of this paper is to revisit the formation of heavy-light molecules under the general
strictures of chiral and heavy quark symmetry, including the mixing between the heavy doublets and their
chiral partners which was partially considered in~\cite{THORSSON,KARLINER,OTHERS,OTHERSX,OTHERSZ}. In leading
order, chiral symmetry  fixes the intra- and cross-multiplet couplings. In particular, 
bound molecules $\bar DD$ with charm and $\bar BB$ with bottom 
may form through channel mixing,
despite the absence of a direct pion coupling by parity.  The P-wave inter-multiplet mixing in the $(0^-,1^-)$ is enhanced by 
the almost degeneracy of the constituents by heavy-quark symmetry, while the S-wave cross-multiplet mixing in the $(0^\pm, 1^\pm)$ is still substantial due to the closeness of the constituents by chiral symmetry.  The latter prevents the
formation of dual chiral molecules such as $\bar {\tilde D}\tilde D$ with charm and $\bar {\tilde B}\tilde B$ with bottom,
as we will show.
Throughout, the coupling to the low-lying 
resonances in the continuum  with more model assumptions 
will be ignored for simplicity. Also interactions mediated by shorter range massive vectors 
and axials will be mostly cutoff through the use of a core cutoff in the pion mediated potential
of $1$ GeV. Only the channels with total angular momenta 
$J\leq 1$ will be discussed.

The organization of the paper is as follows: In section 2 we 
briefly derive the essential contruct for doubly charmed exotic molecules using the strictures of chiral and heavy
quark symmetries and explicit the coupled channel problem for the lowest bound states.  We also 
show how the same coupled channel problem carries to the chiral partners. In section 3, we
extend our analysis to the doubly bottom exotic molecules and their chiral partners. Our conclusions
are given in section 4.

\section{ Charmed exotics molecules}

\subsection{$(0^-,1^-)$ multiplet}

The low energy effective action of heavy-light mesons interacting with pions 
is constrained by both chiral and heavy quark symmetry. In short, the leading part of 
the heavy-light Lagrangian for the charmed multiplet $(0^-, 1^-)$ with pions reads~\cite{MACIEK,ISGUR}

\bea
\label{L-}
{\cal L}\approx &&+2i\left(\bar D \partial_0D+\vec{\bar D}\cdot \partial_0\vec{D}\right)\nonumber\\
&&-\Delta m_{D}\bar DD-\Delta m_{\vec D}\vec{\bar D}\vec{D}\nonumber\\
&&+i\frac{g_H}{f_{\pi}}{\rm Tr}\partial_i\pi \left(D_iD^{\dagger}-DD_i^{\dagger}+\epsilon_{ijk}D_kD_j^{\dagger}\right)
\eea
with $\Delta m_i=m_i-m_C$ of the order of a quark constituent mass. 
The molecular exotics of the type $D\bar D^*$ and alike, follows from (\ref{L-}) through one-pion exchange.
The non-relativistic character of the molecules yield naturally to a Hamiltonian description. 

For all available
2-body channels, the pertinent matrix entries for the interaction are readily found in the form 

\bea
\label{H-}
&&\left<\vec{v_3}\vec{v_4}^{\star}|V|\vec{v_1}\vec{v_2}^{\star}\right>=-C(\vec{v_3}\times\vec{v_1})\cdot \nabla(\vec{v_2}\times\vec{v_4})\cdot \nabla V(r)\nonumber\\
&&\left<0\bar0|V|\vec{v_1}\vec{v_2}^{\star}\right>=C\vec{v_1}\cdot \nabla \vec{v_2}\cdot \nabla V(r)\nonumber\\\
&&\left<\vec{v_2}\bar 0|V|0\vec{v_1}^{\star}\right>=-C\vec{v_1}\cdot \nabla \vec{v_2}\cdot \nabla V(r)\nonumber\\\
&&\left<0\vec{v_3}^{\star}|V|\vec{v_1}\vec{v_2}^{\star}\right>=-C\vec{v_1}\cdot\nabla (\vec{v_2}\times \vec{v_3})\cdot \nabla V(r)\nonumber\\\
&&\left<\bar 0\vec{v_3}|V|\vec{v_1}\vec{v_2}^{\star}\right>=C\vec{v_2}\cdot\nabla (\vec{v_3}\times \vec{v_1})\cdot \nabla V(r)
\eea
 with the isospin factor

\be
C=\vec{I_1}\cdot \vec{I_2}=\left(\left.\frac{1}{4}\right|_{I=1},-\left.\frac{3}{4}\right|_{I=0}\right)
\ee
The spin polarizations of $D^*$ and its conjugate $ \bar{{D}^*}$ are referred to as $\vec v$ and $\vec {v^*}$ respectively. 
Here $V(r)$ is the  regulated one-pion exchange using the standard monopole form factor by analogy with the
pion-nucleon form factor~\cite{BROWN}. 
Denoting by $D_{0\bar 0}(\vec{r})$ the wave function of the molecular scalar, by
$\bar Y_{0\bar i}(\vec r)$ and $Y_{i\bar 0}(\vec r)$ the wavefunctions of the molecular 
vectors, and by $T_{i\bar j}(\vec{r})$ the wavefunction of
the molecular tensors, we can rewrite (\ref{H-}) as

\bea
\label{HWF}
&&(VT)_{k\bar l}=C\epsilon_{kim}\epsilon_{\bar l\bar jn}\partial_{mn}V T_{i\bar j}\nonumber\\
&&(VT)_{0\bar 0}=C\partial_{i\bar j}VT_{i\bar j}\nonumber\\
&&(V\bar Y)_{k\bar 0}=-C\partial_k\partial_{\bar j}V(r)\bar Y_{0\bar j}\nonumber\\
&&(VT)_{0\bar k}=C\epsilon_{\bar k\bar l j}\partial_i\partial_jV(r)T_{i\bar l}\nonumber\\
&&(VT)_{\bar 0 k}=C\epsilon_{ k l j}\partial_{\bar i}\partial_jV(r)T_{l\bar i}
\eea
The explicit reduction of the molecular wavefunctions will be detailed below, for all channels with $J\leq 1$.

\begin{figure}[h!]
  \begin{center}
  \includegraphics[width=6cm]{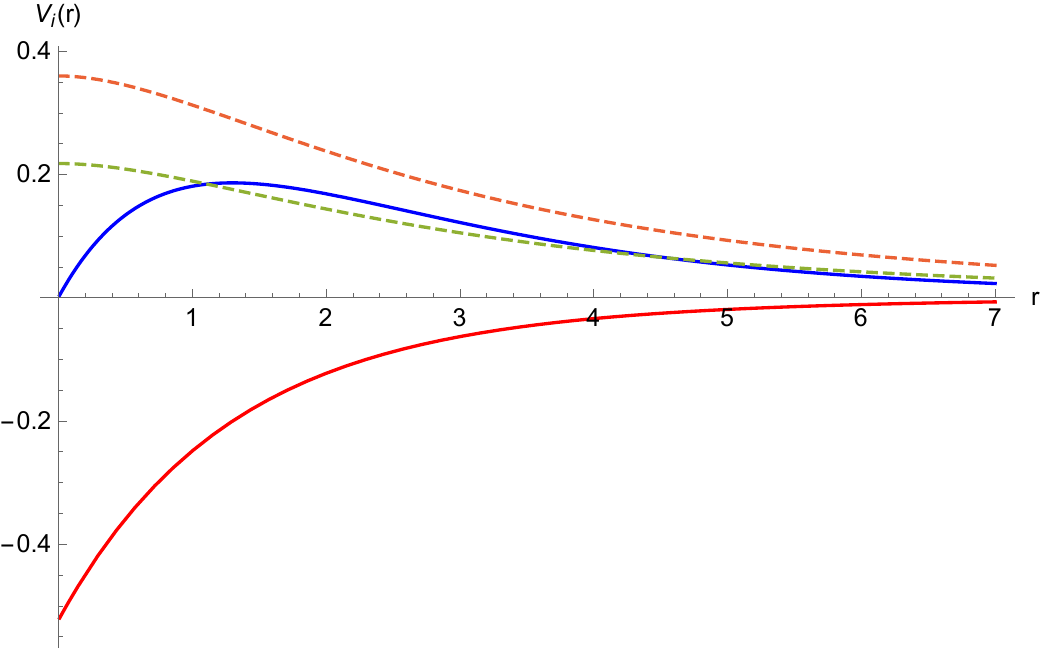}
   \caption{Typical 
   $V_1(r)$ (lower-full-red) and $V_2(r)$ (upper-full-blue) pion induced potentials compared to 
   $\Delta_1^2 V(r)$ (lower-green-dashed) and $\Delta_2^2 V(r)$ (upper-oranged-dashed) to be defined below.}
    \label{fig_v1v2}
  \end{center}
\end{figure}

The one-pion mediated interaction  is defined with 
a core cutoff $\Lambda\gg m_\pi$~\cite{THORSSON,BROWN}

\be
\label{VRR}
&&V(r)=\left(\frac{g_H}{f_\pi}\right)^2\nonumber\\
&&\frac 1{4\pi}\left(\frac{e^{-m_\pi r}}{r}-\frac{e^{-\Lambda r}}{r}-(\Lambda^2-m_\pi^2)\frac{e^{-\Lambda r}}{2\Lambda}\right)
\ee
Once inserted  in (\ref{HWF}) it contributes  a scalar and a  tensor through

\be
\label{VRR}
\partial_i\partial_jV(\vec r)=\delta_{ij}V_1(r)+r_ir_jV_2(r)
\ee
which are shown in Fig.~\ref{fig_v1v2} for $g_H=0.6$~\cite{MACIEK,BARDEEN} and $\Lambda=1$ GeV in units of $\Lambda$. 
The strength of the regulated one-pion exchange potential increases with increasing cutoff $\Lambda$. The dependence of the
results on the choice of core cutoff $\Lambda$ is the major uncertainty of the molecular analysis to follow. The tensor contribution
in (\ref{VRR}) is at the origin of the notorious D-wave admixing in the deuteron state~\cite{BROWN}, and is distinctly different from 
 the gluonic based exchanges in heavy quarkonia~\cite{MANOHAR}.

\subsection{ $(0^+,1^+)$ chiral partners and their mixing}

The leading part of  the heavy-light chiral doublers  Lagrangian for the charmed 
$(0^+, 1^+)$ multiplet with pions reads~\cite{MACIEK}

\bea
\label{L+}
{\cal \tilde L}\approx &&+2i\left(\bar  {\tilde D} \partial_0{\tilde D}+\vec{\bar {\tilde D}}\cdot \partial_0\vec{\tilde D}\right)\nonumber\\
&&-\Delta m_{\tilde D}\bar {\tilde D}\tilde D-\Delta m_{\vec {\tilde D}}\vec{\bar {\tilde D}}\vec{\tilde D}\nonumber\\
&&+i\frac{g_H}{f_{\pi}}{\rm Tr} \partial_i\pi \left(i(\tilde D_i{\tilde D}^{\dagger}+\tilde D\tilde D_i^{\dagger})
+\epsilon_{ijk}\tilde D_k{\tilde D}_j^{\dagger}\right)\nonumber\\
\eea
with again $\Delta m_{\tilde i}=m_{\tilde i}-m_C$ of the order of a quark constituent mass. 
The $(0^+,1^+)$ multiplet mixes with the $(0^-,1^-)$ by chiral symmetry.
The leading part of the interaction in the chirally mixed parity channels reads~\cite{MACIEK,BARDEEN}

\be
\label{L+-}
{\delta \cal L} =\frac{g_{HG}}{f_{\pi}}{\rm Tr}\,\partial_0\pi \left({\tilde D}^{\dagger}_iD_i-i{\tilde D}^{\dagger}D+ {\rm c.c.}\right)
\ee


\subsection{J=0 channels}

To analyze the coupled molecular ground states, we present the analysis for the
$J=0$ coupled channels. We first discuss the mixing in the $(0^-,1^-)$ multiplet, 
followed by the mixing in the $(0^+,1^+)$ chiral mirror multiplet, and finally the 
cross mixing between the $(0^\pm,1^\pm)$ multiplets. The pertinent $0^{PC}$ 
channels  with their spectroscopic ${}^SL_J$ assignments are

\bea
\label{0PC}
0^{++}:&&T^{00}_{i\bar j}({}^1S_0),T^{22}_{ij}({}^5D_0),D_{0\bar 0}^{0}({}^1S^\prime_0)\nonumber\\
0^{-\pm}:&&\bar Y^{1}_{0\bar i}({}^3P_0),Y^{1}_{i\bar0}({}^3P^\prime_0),T^{11}_{i\bar j}({}^3P^{\prime\prime}_0)
\eea
We have added the primes to  track the different contributions in the numerical results below.
Here, $T^{SL,JM}_{ij}$ refers to the tensor spherical harmonics with spin $S$, orbital angular momentum $L$, and total angular momentum $J$ and projection $J_z=M$. As all $JM=00$, we have omitted them in (\ref{0PC})
for convenience.  Also $Y_{\bar 0i}^L\equiv Y^{L,JM}_{i}$
refers to the vector spherical harmonics with orbital angular momentum $L$, 
total angular momentum $J$ with $J_z=M$. 
The explicit form of the  properly normalized tensor and vector spherical harmonics  in this case, 
are readily obtained as

\bea
&&T^{00}_{i\bar j}=\frac{\delta_{i\bar j}}{\sqrt{3}},\qquad 
T^{22}_{i\bar j}=\sqrt{\frac{3}{2}}\left(\hat r_i\hat r_{\bar j}-\frac{\delta_{ij}}{3}\right)\nonumber\\
&&\bar Y^{1}_{0\bar i}=\hat r_{\bar i},\qquad Y^{1}_{i \bar 0}=\hat r_{i},\qquad T^{11}_{i\bar j}=\frac{\epsilon_{i\bar jk}\hat r_k}{\sqrt{2}}
\eea
Here, we note that $T^{00}_{i\bar j}, T^{11}_{i\bar j}, T^{22}_{i\bar j}$ carries explicitly charge conjugation
$C=+$. However, $\bar Y^1_{0\bar k},Y^1_{k\bar 0}$  carry $C=\pm$. It is
straightforward to project onto states of good $C$ and rewrite the interactions to follow in this basis, 
but for $J=0$ it is not needed, as only the $C=+$ combination is seen not to vanish. It will not be the case
for $J=1$ as we will discuss below. 
The even-parity channels $T^{0},T^{2},D^{0}$ mix
, and the odd-parity channels $T^{1},Y^{1},\bar Y^{1}$ mix. 

\subsection{Interaction in $(0^-,1^-)$ multiplet: $J=0$}

The mixing part of the interaction in the $J^{PC}=0^{++}$ channel is

\be
\label{H126}
V^{0++}=C\left(\begin{array}{ccc}
2V_1+\frac{2V_2}{3}&-\frac{\sqrt{2}V_2}{3}&\sqrt{3}V_1+\frac{V_2}{\sqrt{3}}\\
-\frac{\sqrt{2}V_2}{3}&-V_1+\frac{V_2}{3}&\sqrt{\frac{2}{3}}V_2\\
\sqrt{3}V_1+\frac{V_2}{\sqrt{3}}&\sqrt{\frac{2}{3}}V_2&0\\
\end{array}\right)
\ee
while in the  $J^{PC}=0^{-\pm}$ channel it is 

\be
\label{H345}
V^{0-\pm}=C\left(\begin{array}{ccc}
0&-V_1-V_2&-\sqrt{2}V_1\\
-V_1-V_2&0&\sqrt{2}V_1\\
-\sqrt{2}V_1&\sqrt{2}V_1&V_1+V_2\\
\end{array}\right)
\ee
We note that while the one-pion mediated $D\bar D\rightarrow D\bar D$  intraction in (\ref{H126}) vanishes by parity, 
the cross interactions $D\bar D\rightarrow D^* \bar{D^*}$ and ${D^*}\bar{D}\rightarrow D\bar{D^*}$ do not.
As a result bound states through mixing $D\bar D\rightarrow D^* \bar{D^*} \rightarrow D\bar D$ could 
and will form
in the same order. The corresponding mass shifts and  kinetic terms are

\bea
\label{K126}
&&K^{0++}=\nonumber\\
&&\left(\begin{array}{ccc}
4m_1-\frac{\nabla_r^2}{2m_1}&0&0\\
&4m_1-\frac{\nabla_r^2}{2m_1}+\frac 3{m_1r^2}&0\\
0&0&4m_2-\frac{\nabla_r^2}{2m_2}\\
\end{array}\right)\nonumber\\
\eea
and

\tiny
\bea
\label{K345}
&&K^{0-\pm}=\nonumber\\
&&\left(\begin{array}{ccc}
m_{12}-\frac{\nabla_r^2}{2m_3}+\frac{1}{m_3r^2}&0&0\\
0&2m_{12}-\frac{\nabla_r^2}{2m_3}+\frac{1}{m_3r^2}&0\\
0&0&4m_1-\frac{\nabla_r^2}{2m_1}+\frac{1}{m_1r^2}\\
\end{array}\right)\nonumber\\
\eea
\normalsize
with $\nabla_r^2\phi=\frac{1}{r}\frac{d^2}{dr^2}(r\phi)$, $m_{12}=m_1+m_2$ 
and the corresponding reduced masses are 

\be
\label{M123}
\left(m_1\equiv\frac{m_{\vec{D}}}{2},\,\,\,m_2\equiv\frac{m_D}{2},\,\,\,m_3\equiv\frac{m_Dm_{\vec{D}}}{m_D+m_{\vec{D}}}\right)
\ee
The empirical masses are
$m_{D^\pm}=1.870$ GeV, $m_{D^0}=1.865$ GeV and $m_{D^{*\pm}}=2.010$ GeV, $m_{D^{*0}}=2.007$ GeV.
Below, we will use the averages over the isotriplets for $m_{1,2,3}$. Specifically, $m_1=1.005$ GeV, $m_2=0.934$ GeV and $m_3=0.968$ GeV. 
Here $m_\pi=137$ MeV and $f_\pi=93$ MeV, with $g_H=0.6$~\cite{MACIEK,BARDEEN}.  




\subsection{Interaction in $(0^+,1^+)$ multiplet: $J=0$}

For the $(0^+,1^+)$ multiplet, the classification of all the states remains the same.  The relation
between the matrix elements in the $(0^-,1^-)$ sector and the $(0^+,1^+)$ sector (primed below) 
can be made  explicit if we note the relations

\bea
\label{T1}
&&\left.|0\right>^{\prime }= -i\left.|0\right>\nonumber\\
&&\left.|\bar 0\right>^{\prime }=+i\left.|0\right>
\eea
With this in mind, the matrix elements  between the different tensor projections are related as follows

\bea
\label{T2}
\left<T|H|T\right>^{\prime}=&&+\left<T|H|T\right>\nonumber\\
\left<\bar Y|H|T\right>^{\prime}=&&+i\left<\bar Y|H|T\right>\nonumber\\
\left<Y|H|T\right>^{\prime}=&&-i\left<Y|H|T\right>\nonumber\\
\left<\bar Y|H|Y\right>^{\prime }=&&-\left<\bar Y|H|Y\right>
\eea
As a result we have $\tilde V^{0++}=V^{0++}$ and

\bea
\tilde V^{0-\pm}=\label{H345}
C\left(\begin{array}{ccc}
0&V_1+V_2&-i\sqrt{2}V_1\\
V_1+V_2&0&-i\sqrt{2}V_1\\
i\sqrt{2}V_1&i\sqrt{2}V_1&V_1+V_2\\
\end{array}\right)
\eea
The kinetic contributions $\tilde K^{0++}$ and $\tilde K^{0++}$ follows from (\ref{K126}-\ref{K345}) with the appropriate substitution for the reduced masses.  We will use the empirical masses for the reported chargeless doublet $(D_0^*, D_1^0)$ with 
$m_{\tilde{D}}=2.400$ GeV and $m_{{\tilde{D}}^*}=2.420$ GeV, which translate to
$\tilde m_1=1.210$ GeV, $\tilde m_2=1.200$ GeV and $\tilde m_3=1.205$ GeV. Here
$\tilde g_H=0.6$ follows from heavy quark symmetry.

\subsection{Interaction across $(0^\pm ,1^\pm)$ multiplets: $J=0$}

The mixed coupling between the $(0^-,1^-)$ and $(0^+,1^+)$ induces a scalar interaction typically of the form 
$\delta V(r)\approx \Delta m^2V(r)$ with $\Delta m/m_1\approx 0.4/1.2=1/3$.  
In the relevant range shown in  Fig.~\ref{fig_v1v2}, it is about the same as $V_1(r)$
and will be retained. The  corresponding
one-pion mediated potential  in the $0^{++}$ is

\bea
W^{0++}=\label{V126+-}
\left(\frac{g_{GH}}{g_H}\right)^2CV
\left(\begin{array}{ccc}
-\Delta_1^2&0&0\\
0&-\Delta_1^2&0\\
0&0&-\Delta_2^2
\end{array}\right)
\eea
and in the $0^{-\pm}$  channel,  is

\bea
\label{V345+-}
W^{0-\pm}=\left(\frac{g_{GH}}{g_H}\right)^2CV
\left(\begin{array}{ccc}
i\Delta_1\Delta_2&0&0\\
0&-i\Delta_1\Delta_2&0\\
0&0&-\Delta_1^2
\end{array}\right)
\eea
Here the empirical mass splittings are

\bea
&&\Delta_1=(m_{\tilde D^{\star}}-m_{D^{\star}})\approx 410\, {\rm MeV}\nonumber\\
&&\Delta_2=(m_{\tilde D}-m_{D})\approx 530\, {\rm MeV}
\eea

The stationary coupled channel problem for the ground states in $J^{PC}=0^{++}$ and $J^{PC}=0^{--}$,  follows from 
the $6\times 6$ eigenvalue problem $\left({\bf H}={\bf K}+{\bf V}+{\bf  W}\right)\Phi_i=E\Phi_i$ with now $\Phi_i\equiv r\phi_i$. 
To proceed further, we need to solve the coupled channels problem numerically with 

\bea
{\bf H}=\left(\begin{array}{cc}
K+V&W^{\dagger}\\
W&\tilde K+\tilde V
\end{array} \right)
\eea
in each sector.

\subsection{$J=1$ channels}

The pertinent projections onto the higher $J^{PC}$ channels of the molecular wavefunctions in (\ref{HWF})
require the use of both vector and higher tensor spherical harmonics~\cite{EDMONDS,THORN}. For $J=1$,
we will use the explicit forms quoted in~\cite{THORN} with the ${}^SL_J$ 
assignment completly specified. For the $(1^\mp,0^\mp)$ multiplets, there are 4 different $1^{PC}$ sectors

\bea
1^{++}:&&T^{2,2}_{i\bar j}({}^5D_1),Y^{0+}_{i}({}^3S_1),Y^{2+}_{i}({}^3D_1)\nonumber\\
1^{--}:&&T^{0,1}_{i\bar j}({}^1P_1),T^{2,1}_{i\bar j}({}^5P_1),T^{2,3}_{i\bar j}({}^5F_1),\nonumber\\
&&Y^{1-}_{i}({}^3P_1),D^{1}({}^1P^\prime_1)\nonumber\\
1^{+-}:&&T^{1,0}_{i\bar j}({}^3S_1),T^{1,2}_{i\bar j}({}^3D_1),Y^{0-}_{i}({}^3S^\prime_1),Y^{2-}_{i}({}^3D^\prime_1)\nonumber\\
1^{-+}:&&T^{1,1}_{i\bar j}({}^3P_1),Y^{1+}_{i}({}^3P^\prime_1)
\eea
with the $JM$ labels omitted for convenience.
The normalized tensor harmonics are~\cite{THORN}

\be
\label{J1LSJ}
&&T^{01,1m}_{ij}=\frac{\delta_{ij}}{\sqrt{3}}Y_{1m}\nonumber\\
&&T^{21,1m}_{ij}=\sqrt{\frac{3}{5}}\left(\frac{\delta_{ij}}{3}-\hat r_i\hat r_j\right)Y_{1m}\nonumber \\ 
&&\qquad\qquad-\sqrt{\frac{3}{10}}(r_i\nabla_j+r_j\nabla_i)Y_{1m}\nonumber\\
&&T^{23,1m}_{ij}=3\sqrt{\frac{1}{10}}\left(\frac{\delta_{ij}}{3}-\hat r_i\hat r_j\right)Y_{1m}\nonumber \\ 
&&\qquad\qquad-\sqrt{\frac{1}{5}}(r_i\nabla_j+r_j\nabla_i)Y_{1m}\nonumber\\
&&T^{22,1m}_{ij}=\frac{1}{2}(r_iL_j+r_jL_i)Y_{1m}\nonumber\\
&&Y^{0,1m}_{i}=\frac{1}{\sqrt{3}}({\sqrt{2}}r\nabla_iY_{1m}+\hat r_iY_{1m})\nonumber\\
&&Y^{2,1m}_{i}=\frac{1}{\sqrt{3}}(r\nabla_iY_{1m}-\sqrt{2}\hat r_iY_{1m})\nonumber\\
&&Y^{1,1m}_{i}=\frac{1}{\sqrt{2}}iL_{i}Y_{1m}
\ee
The $DD^*$ channels  with definite charge conjugation 
$C=\pm$  are explicitly 

\bea
\label{CPSIGNS}
&&Y_{i}^{0\pm}=\frac{1}{\sqrt{2}}(\bar Y^{0}_{0\bar i}\pm Y^{0}_{i\bar 0})\nonumber\\
&&Y_{i}^{2\pm}=\frac{1}{\sqrt{2}}(\bar Y^{2}_{0\bar i}\pm Y^{2}_{i\bar 0})\nonumber\\
&&Y_{i}^{1\pm}=\frac{1}{\sqrt{2}}(\bar Y^{1}_{0\bar i}\mp Y^{1}_{i\bar 0})
\eea
We note that 

\be
T^{1L,JM}_{i\bar j}=\frac{\epsilon_{i\bar jk}}{\sqrt{2}}Y^{L,JM}_{k}
\ee

\subsection{Interaction in $(0^-,1^-)$ multiplet: $J=1$}

In terms of the previous $J=1$ channels, the one-pion mediated interaction in the $(0^-,1^-)$ multiplet
in the $J^{PC}=1^{--}$ channel  takes the block form

\bea
V^{1--}=C\left(\begin{array}{cc}
v_0&v_1\\
v_1^{\dagger}&v_2
\end{array}\right)
\eea
with the blocks defined as

\bea
v_0=\left(\begin{array}{ccc}
2V_1+\frac{2V_2}{3}&\frac{2V_2}{3\sqrt{5}}&-\sqrt{\frac{2}{15}}V_2\\
\frac{2V_2}{3\sqrt{5}}&-V_1+\frac{2V_2}{15}&-\frac{\sqrt{6}}{15}V_2\\
-\sqrt{\frac{2}{15}}V_2&-\frac{\sqrt{6}}{15}V_2&-V_1+\frac{V_2}{15}
\end{array}\right)
\eea

\bea
v_1=\left(\begin{array}{ccc}
0&\sqrt{3}V_1+\frac{V_2}{\sqrt{3}}\\
\frac{\sqrt{3}V_2}{\sqrt{5}}&-\frac{2V_2}{\sqrt{15}}\\
\frac{\sqrt{2}V_2}{\sqrt{5}}&\sqrt{\frac{2}{5}}V_2
\end{array}\right)
\eea

\bea
v_2=\left(\begin{array}{cc}
-V_1&0\\
0&0
\end{array}\right)
\eea

Similarly, the one-pion mediated interaction in the $J^{PC}=1^{+-}$ channel has the following block structure

\bea
V^{1+-}=C\left(\begin{array}{cc}
v_3&v_4\\
v_4^{\dagger}&v_5
\end{array}\right)
\eea
with each block defined as

\bea
v_3=v_5=\left(\begin{array}{cc}
V_1+\frac{V_2}{3}&-\frac{\sqrt{2}V_2}{3}\\
-\frac{\sqrt{V_2}}{3}&V_1+\frac{2V_2}{3}
\end{array}\right)
\eea

\bea
v_4=\left(\begin{array}{cc}
-2V_1-\frac{2V_2}{3}&-\frac{\sqrt{2}V_2}{3}\\
-\frac{\sqrt{2}V_2}{3}&-2V_1-\frac{V_2}{3}\\
\end{array}\right)
\eea

In the remaining $J^{PC}=1^{-+}$ and $J^{PC}=1^{++}$ the one-pion mediated interactions are respectively given by

\bea
V^{1-+}=C\left(\begin{array}{cc}
V_1&-2V_1-V_2\\
-2V_1-V_2&V_1
\end{array}\right)
\eea
and

\bea
V^{1++}=C\left(\begin{array}{ccc}
-V_1&i\frac{\sqrt{2}V_2}{\sqrt{3}}&\frac{iV_2}{\sqrt{3}}\\
-i\frac{\sqrt{2}V_2}{\sqrt{3}}&-\frac{3V_1+V_2}{3}&\frac{\sqrt{2}V_2}{3}\\
-\frac{iV_2}{\sqrt{3}}&\frac{\sqrt{2}V_2}{3}&-\frac{3V_1+2V_2}{3}
\end{array}\right)
\eea

 \begin{figure}[h!]
 \begin{center}
  \includegraphics[width=6cm]{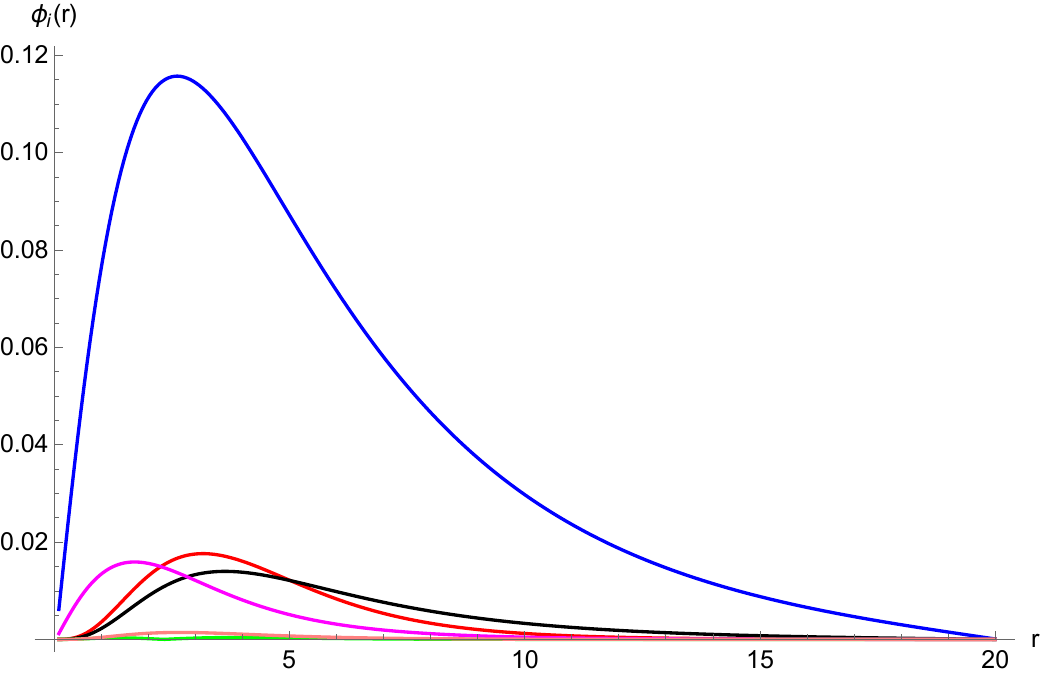}
   \includegraphics[width=6cm]{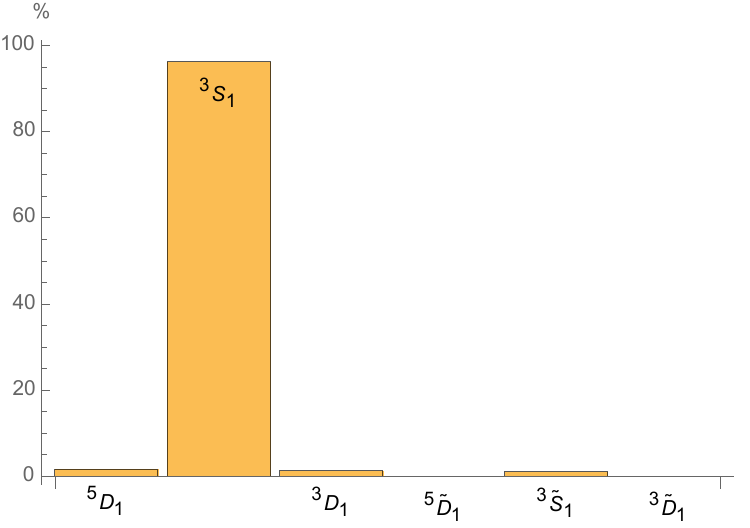}
   \caption{Typical $\Phi$ radial 
   wavefunctions in the mixed $J^{PC}=1^{++}$ channel (upper plot), 
   for the lowest bound state for the charm exotic state with $C=-3/4$ (isosinglet), in units of $\Lambda=1$ GeV. 
   The corresponding percentages content of the bound wavefunction in the $J^{PC}=1^{++}$ channel are shown below,
   with their spectroscopic assignments.}
    \label{fig_fcj1-1++}
  \end{center}
\end{figure}

\begin{figure}[h!]
 \begin{center}
  \includegraphics[width=6cm]{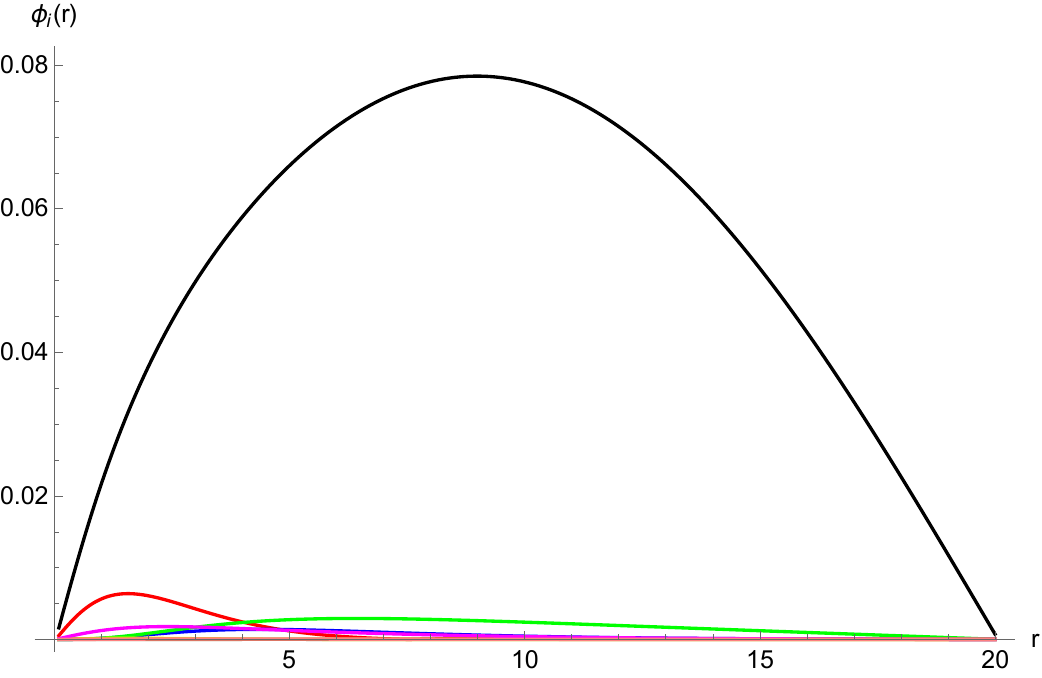}
   \includegraphics[width=6cm]{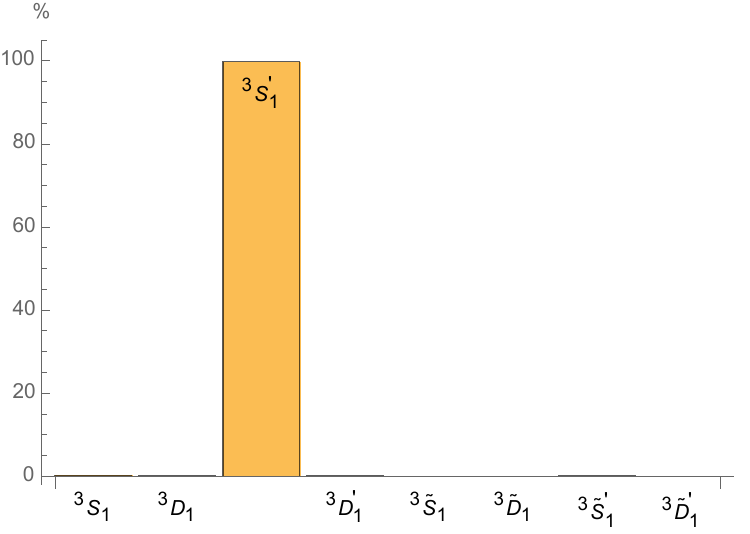}
   \caption{Typical $\Phi$ radial 
   wavefunctions in the mixed $J^{PC}=1^{+-}$ channel (upper plot), 
   for the lowest {\bf unbound} state for the charm exotic state with $C=1/4$ (isotriplet), in units of $\Lambda=1$ GeV. 
   The corresponding percentages content of the {\bf unbound}  wavefunction in the $J^{PC}=1^{+-}$ channel are shown below,
   with their spectroscopic assignments.}
    \label{fig_fcj1-1+-}
  \end{center}
\end{figure}

\subsection{Interaction across $(0^\pm,1^\pm)$ multiplets: $J=1$}

The one-pion mediated interaction within the $(0^+, 1^+)$ multiplet follows the same construct as
in the $(0^-,1^-)$ multiplet using the transfer rules in (\ref{T1}-\ref{T2}). The same interaction across
the two chiral multiplets introduces also a diagonal mixing of the form

\bea
\left<T|V|\tilde T\right>=&&-CV\left(\frac{g_{GH}}{g_{H}}\right)^2\Delta_1^2\,\delta_{T \tilde T}\nonumber\\
\left<D|V|\tilde D\right>=&&-CV\left(\frac{g_{GH}}{g_{H}}\right)^2\Delta_2^2\,\delta_{D \tilde D}\nonumber\\
\left<Y|V|\tilde Y\right>=&&-iCV\left(\frac{g_{GH}}{g_{H}}\right)^2\Delta_1 \Delta_2\,\delta_{Y \tilde Y}
\eea
The $\tilde D \tilde D^{\star}$ states with good charge conjugation follows from (\ref{CPSIGNS})  through the
substitution $\pm\rightarrow \mp$ only on the right hand side. The total Hamiltonian in the $(0^+, 1^+)$ sector 
to diagonalize is $H=K+V$. Including the chiral multiplet, the total Hamiltonian across the  $(0^\pm, 1^\pm)$ sectors
to diagonalize is twice larger  ${\bf H}={\bf K}+{\bf V}+{\bf W}$.

\subsection{Results for charm exotic molecules}

In the upper plot of Fig.~\ref{fig_fcj1-1++} we show the typical $\Phi_i$  radial components of the bound 
isosinglet charm wavefunction 
with energy $E=3.867$ GeV  for a cutoff $\Lambda=1$ GeV,
as a function of the radial distance $r$ also in units of 1 GeV.  The chiral cross coupling between the $(0^-,1^-)$ and $(0^+,1^+)$ multiplets
induces a very small mixing to the molecular wavefunction in the $(0^-,1^-)$ multiplet as displayed  in Fig.~\ref{fig_fcj1-1++}.  In the
lower chart of Fig.~\ref{fig_fcj1-1++} we show the percentage content of the contributions to the same wavefunction,
with the ${}^SL_J$ assignments referring to the $(0^-,1^-)$ multiplet, 
and the ${}^S\tilde L_J$ assignments referring to the $(0^+,1^+)$ multiplet.
The mixing results in a stronger binding in this channel wich is mostly an isosinglet ${}^1S_3$ contribution in the 
($1^-,0^-$) multiplet with almost no D-wave admixture. This molecular state carries $J^{PC}=1^{++}$ assignment, and from
our ${}^SL_J$ assignments in (\ref{J1LSJ}) it is chiefly  an isosinglet  $D\bar D^*$ molecule. We identify this state with the reported
isosinglet exotic $X(3872)$. The cross chiral mixing causes the dual chiral partners  $\tilde D \bar{\tilde D}^*$ state to unbind. .

In the upper plot of Fig.~\ref{fig_fcj1-1+-} we show the typical $\Phi_i$  radial components of the {\bf unbound}
isotriplet charm wavefunction for a cutoff $\Lambda=1$ GeV,
as a function of the radial distance $r$ also in units of 1 GeV.  The multi-channel coupling  in the channel with $J^{PC}=1^{+-}$, 
shows that the dominant wave is ${}^3S^\prime_1$ which is composed of a resonating isosinglet $D\bar D^*$(3876) compound. 
The wave shows a weak visible attraction near the origin that is not enough to bind. We conclude that the reported 
$Z_c(3900)^\pm$ is at best a resonance in the continuum in our analysis. All other
channels are unbound for  charm exotic molecules with our  cutoff of 1 GeV, in both the isosinglet and isotriplet configurations

\begin{figure}[h!]
 \begin{center}
  \includegraphics[width=6cm]{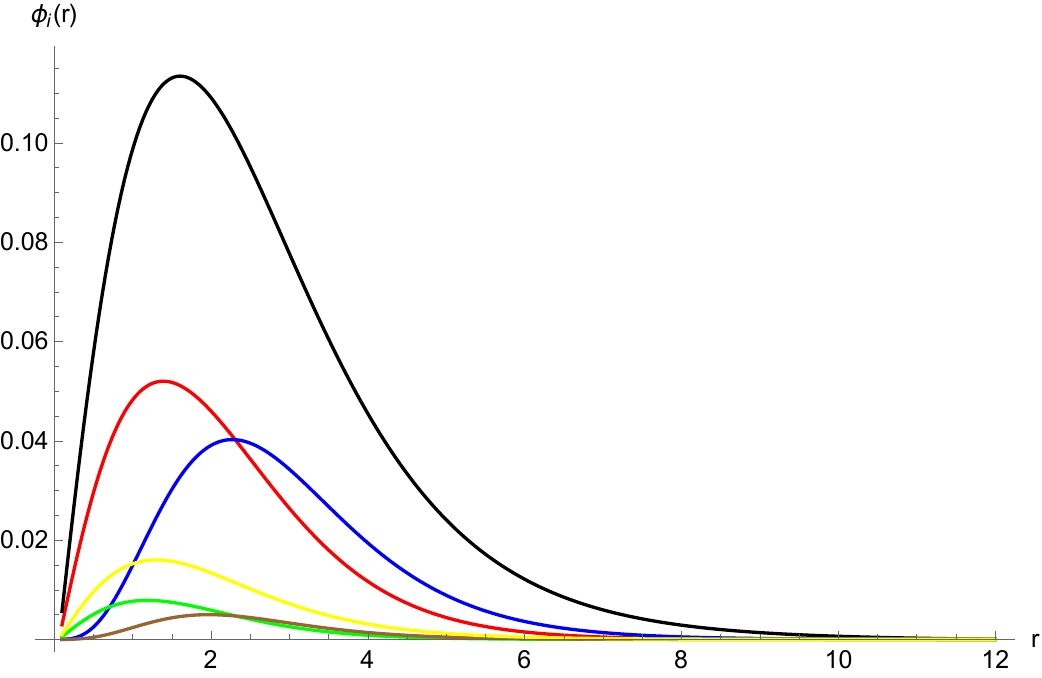}
   \includegraphics[width=6cm]{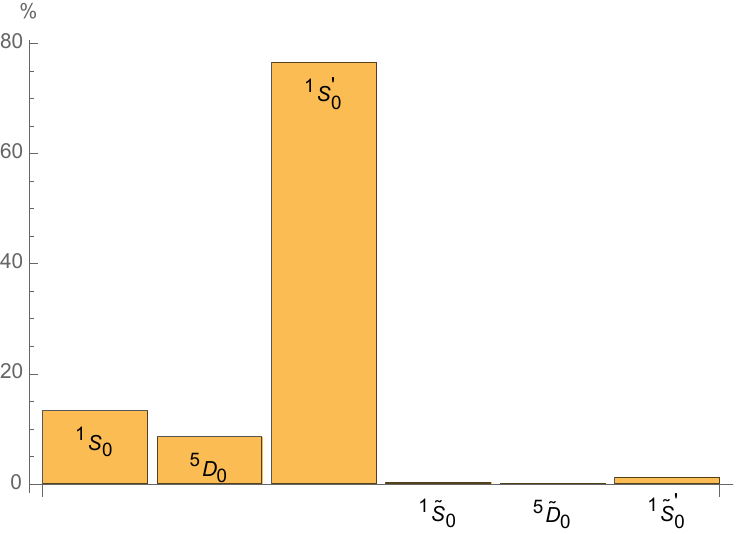}
   \caption{Typical $\Phi$ radial 
   wavefunctions in the mixed $J^{PC}=0^{++}$ channel (upper plot), 
   for the lowest bound state for the bottom exotic state with $C=-3/4$ (isosinglet), in units of $\Lambda=1$ GeV. 
   The corresponding percentages content of the bound wavefunction in the $J^{PC}=0^{++}$ channel are shown below, with their spectroscopic assignments.}
   \label{fig_fbj0-0++}
  \end{center}
\end{figure}

 \begin{figure}[h!]
 \begin{center}
  \includegraphics[width=6cm]{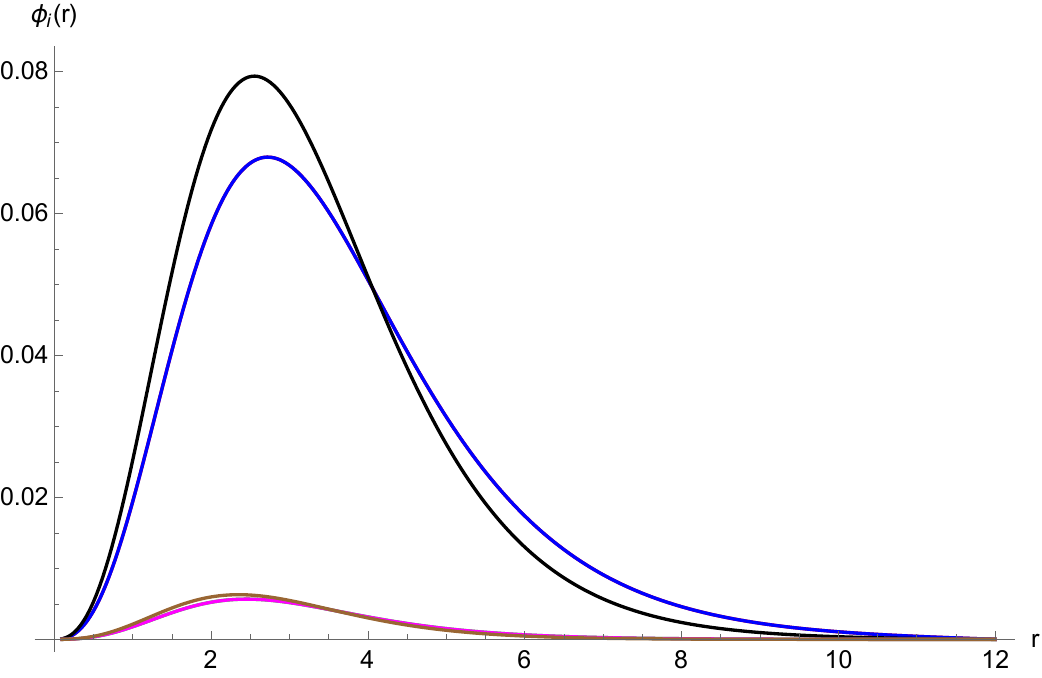}
   \includegraphics[width=6cm]{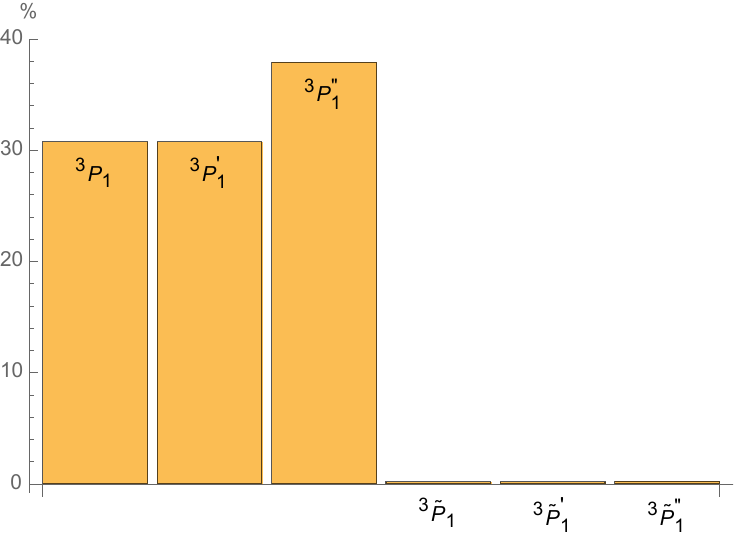}
   \caption{Typical $\Phi$ radial 
   wavefunctions in the mixed $J^{PC}=0^{-+}$ channel (upper plot), 
   for the lowest bound state for the bottom exotic state with $C=-3/4$ (isosinglet), in units of $\Lambda=1$ GeV. 
   The corresponding percentages content of the bound wavefunction in the $J^{PC}=0^{-+}$ channel are shown below, with their spectroscopic assignments.}
   \label{fig_fbj0-0-+}
  \end{center}
\end{figure}

\section{Bottom exotic molecules}

Doubly bottom exotic molecules follow the same construction as before, all potentials and inteactions remains of the
the same form , with now the new mass parameters

\be
\left(m_1\equiv\frac{m_{\vec{B}}}{2},\,\,\,m_2\equiv\frac{m_B}{2},\,\,\,m_3\equiv\frac{m_Bm_{\vec{B}}}{m_B+m_{\vec{B}}}\right)
\ee
For the $(0^-, 1^-)$ multiplet, we have $m_B=5.279$ GeV, $m_{\vec B}=5.325$ GeV and $m_{\tilde B}\approx m_{\tilde{\vec B}}=5.727$ GeV, and therefore
 $m_1=2.662$ GeV, $m_2=2.640$ GeV and $m_3=2.651$ GeV.   For the $(0^+, 1^+)$ multiplet, we have $m_{\tilde{\vec B}}=5.727$ GeV. Assuming a common splitting 
 $m_{\vec B}-m_B=m_{\tilde{\vec B}}-m_{\tilde B}=46$ MeV, we have $m_{\tilde B}=5.681$ GeV and therefore
 $\tilde m_1=2.869$ GeV, $\tilde m_2=2.840$ GeV and $\tilde m_3=2.852$ GeV.
  The results for the chirally mixed states for the bottom exotic states involving the pair multiplet $(0^\pm, 1^\pm)$ 
 can be obtained using similar arguments to those used for charm with the same cutoff choice. Since the one-pion
 exchange interaction is three times stronger in the isosinglet channel  than the isotriplet channel, a multitude of
 isosinglet bottom exotic states will be revealed, thanks also to the heavier bottom mass and thus smaller kinetic energy
 in comparison to  the charm exotic states.

 \begin{figure}[h!]
 \begin{center}
  \includegraphics[width=6cm]{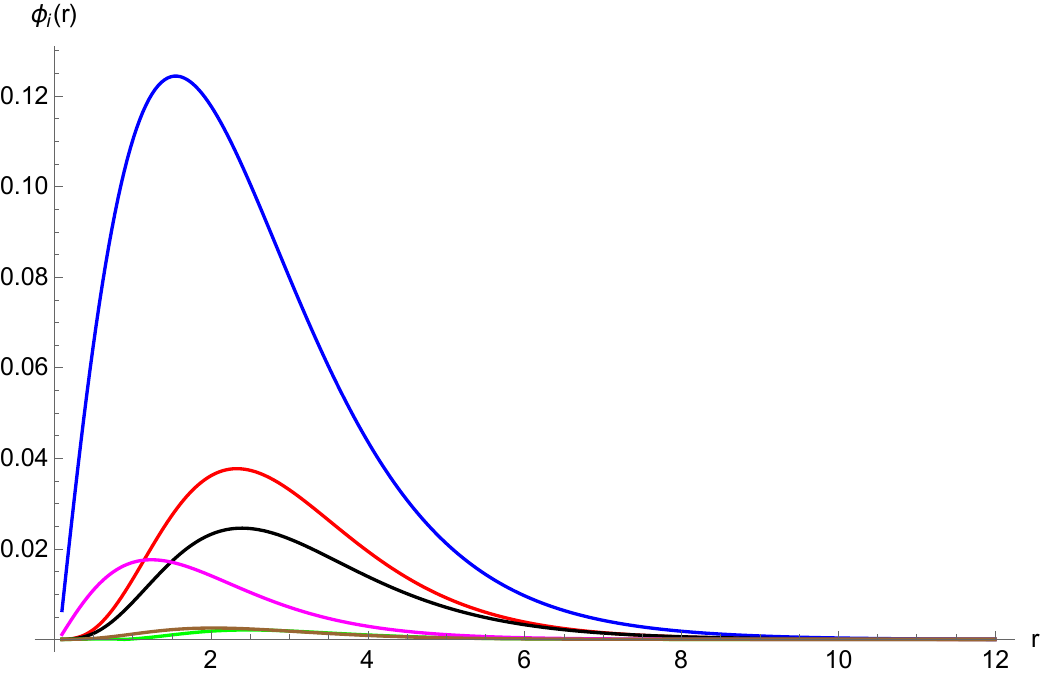}
   \includegraphics[width=6cm]{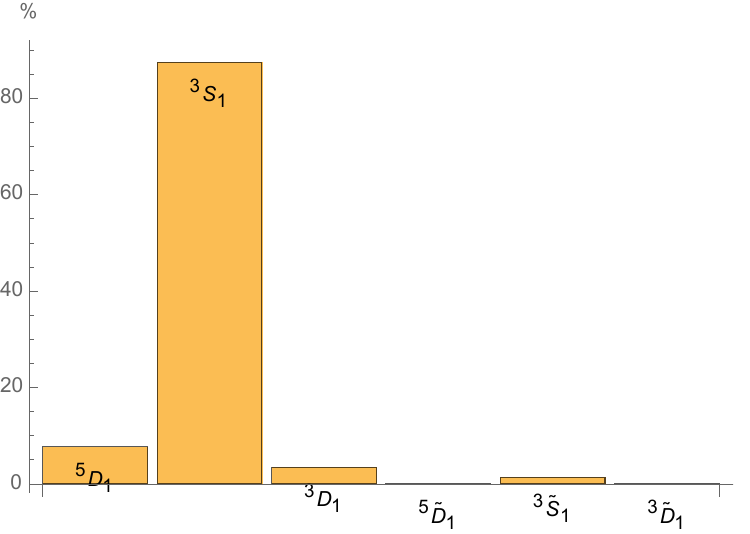}
   \caption{Typical $\Phi$ radial 
   wavefunctions in the mixed $J^{PC}=1^{++}$ channel (upper plot), 
   for the lowest bound state for the bottom exotic state with $C=-3/4$ (isosinglet), in units of $\Lambda=1$ GeV. 
   The corresponding percentages content of the bound wavefunction in the $J^{PC}=1^{++}$ channel are  shown below, with their spectroscopic assignments.}
   \label{fig_fbj1-1++}
  \end{center}
\end{figure}

 \begin{figure}[h!]
 \begin{center}
  \includegraphics[width=6cm]{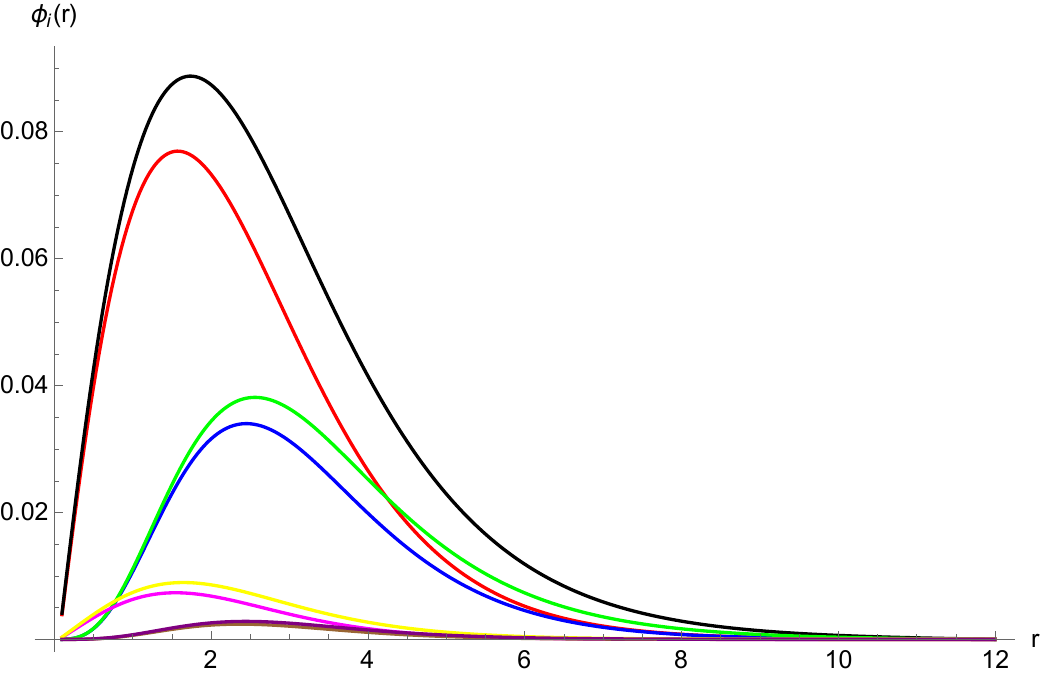}
   \includegraphics[width=6cm]{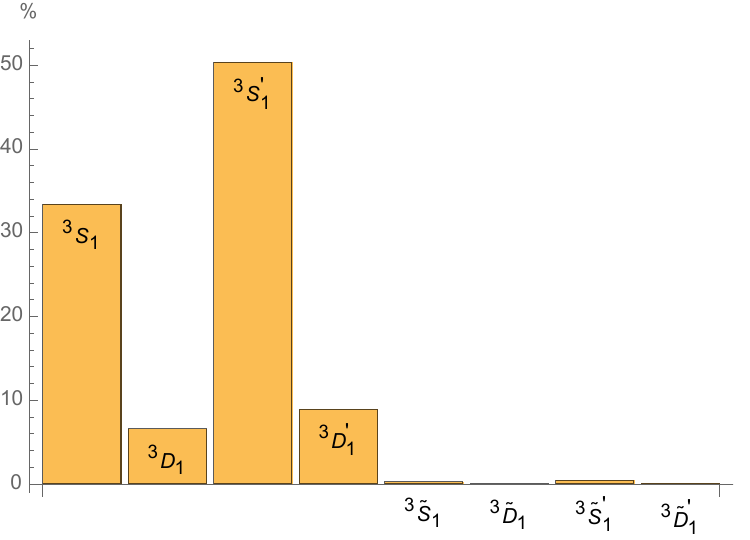}
   \caption{Typical $\Phi$ radial 
   wavefunctions in the mixed $J^{PC}=1^{+-}$ channel (upper plot), 
   for the lowest bound state for the bottom exotic state with $C=-3/4$ (isosinglet), in units of $\Lambda=1$ GeV. 
   The corresponding percentages content of the bound wavefunction in the $J^{PC}=1^{+-}$ channel are shown below, with their spectroscopic assignments.}
   \label{fig_fbj1-1+-}
  \end{center}
\end{figure}

 \subsection{Results for bottom exotic molecules}
 
 In Fig.~\ref{fig_fbj0-0++} we show the behavior of the typical isosinglet bound state wavefunctions
contributing in the $J^{PC}=0^{++}$ channel with energy $E=$ 10.509 GeV ($\Lambda=1$ GeV)
(upper-plots). The percentage content of the same wavefunction is displayed as a histogram with the appropriate 
parity labels in the lower display, with the ${}^SL_J$ assignments referring to the $(0^-,1^-)$ multiplet, 
and the ${}^S\tilde L_J$ assignments referring to the $(0^+,1^+)$ multiplet. From the assignments given in (\ref{0PC}),
we see that the $0^{++}$ mixed bound state is chiefly an isosinglet $B\bar B$ (${}^1S_0$) molecule, with  relatively small 
$B^*\bar B^*$ (${}^1S_0$ ) and $B^*\bar B^*$ (${}^5D_0$ ) admixtures.

 In Fig.~\ref{fig_fbj0-0-+} we show the behavior of the typical isosinglet bound state wavefunctions
contributing in the $J^{PC}=0^{-+}$ channel with energy $E=$10.555 GeV ($\Lambda=1 $ GeV)
(upper-plots). The percentage content of the same wavefunction is displayed as a histogram with the appropriate 
parity labels in the lower display, with the ${}^SL_J$ assignments referring to the $(0^-,1^-)$ multiplet, 
and the ${}^S\tilde L_J$ assignments referring to the $(0^+,1^+)$ multiplet. From the assignments given in (\ref{0PC}),
we see that the $0^{-+}$ mixed bound state is  a mixed molecule with about equal admixture of
$B\bar B^*$ (${}^3P_0$ ), $B^*\bar B$ (${}^3P^\prime_0$ ) and $B^*\bar B^*$ (${}^3P^{\prime\prime}_0$) molecules all
from the $(0^-,1^-)$ multiplet as those from the $(0^+,1^+)$ are shown to decouple and unbind. The
effect of the latters is to cause the formers to bind twice more.

In Fig.~\ref{fig_fbj1-1++} we show the behavior of the typical isosinglet bound wavefunctions
contributing in the $J^{PC}=1^{++}$ channel with energy $E=10.532$ GeV ($\Lambda=1 $ GeV)
(upper-plots). The percentage content of the same wavefunction is displayed as a histogram with the appropriate 
parity labels in the lower display, with the ${}^SL_J$ assignments referring to the $(0^-,1^-)$ multiplet, 
and the ${}^S\tilde L_J$ assignments referring to the $(0^+,1^+)$ multiplet. From the assignments given in (\ref{J1LSJ}),
we see that the $1^{++}$ mixed bound state is chiefly a $B\bar B^*$ (${}^3S_1$), with small 
$B^*\bar B^*$ (${}^5D_1$ ) and $B\bar B^*$ (${}^3D_1$) admixtures. We see again the decoupling of the 
molecular configurations with ${}^S\tilde L_J$ assignments as they are found to unbind, leaving 
 the ${}^SL_J$ assignments twice more bound as per our calculation. A quick comparison  between 
 Fig.~\ref{fig_fcj1-1++} and  Fig.~\ref{fig_fbj1-1++} shows that this neutral bottom molecular state is the mirror
 analogue of the neutral charm molecular state which we suggest  as $X_b(10532)$.

In Fig.~\ref{fig_fbj1-1+-} we show the behavior of the typical isosinglet bound wavefunctions
contributing in the $J^{PC}=1^{+-}$ channel with energy $E=10.550$ GeV ($\Lambda=1 $ GeV)
(upper-plots). The percentage content of the same wavefunction is displayed as a histogram with the appropriate 
parity labels in the lower display, with the ${}^SL_J$ assignments referring to the $(0^-,1^-)$ multiplet, 
and the ${}^S\tilde L_J$ assignments referring to the $(0^+,1^+)$ multiplet. From the assignments given in (\ref{J1LSJ}),
we see that the $1^{+-}$ mixed bound state are primarily $B^*\bar B^*$ (${}^3S_1$) and 
$B\bar B^*$ (${}^3S^\prime_1$) molecules, with small $B^*\bar B^*$ (${}^3D_1$) and 
$B\bar B^*$ (${}^3D^\prime_1$) molecular admixtures. 
The molecules are mostly from the $(0^-,1^-)$ multiplet as those from the $(0^+,1^+)$ are shown to decouple and unbind. Again,
the effect of the latters is to cause the formers to bind twice more.

 \begin{figure}[h!]
 \begin{center}
  \includegraphics[width=6cm]{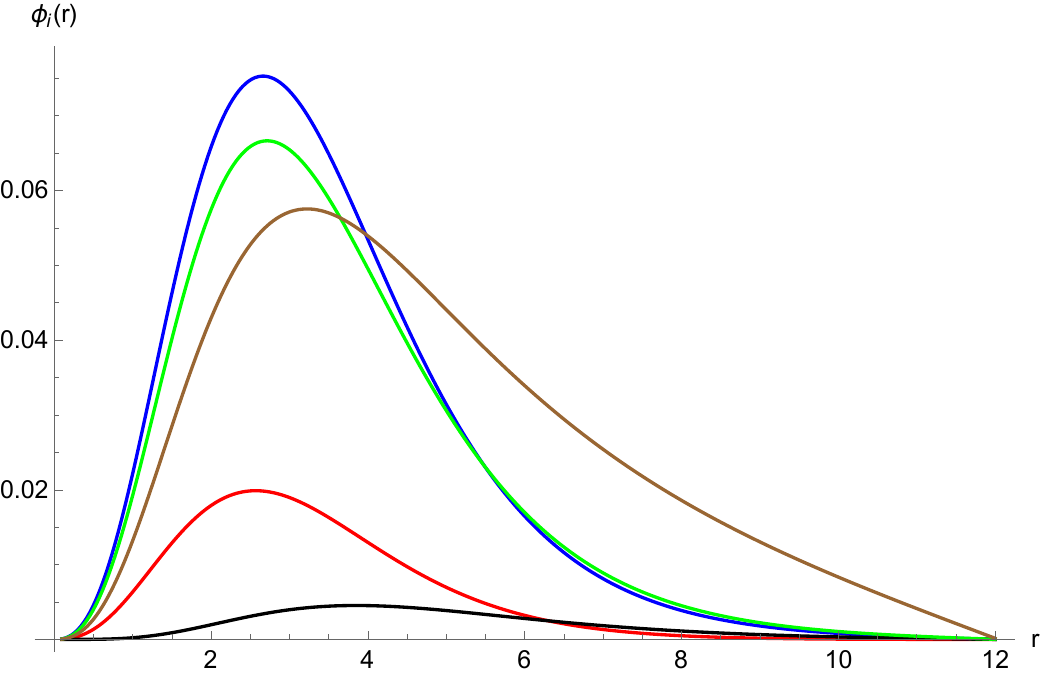}
   \includegraphics[width=6cm]{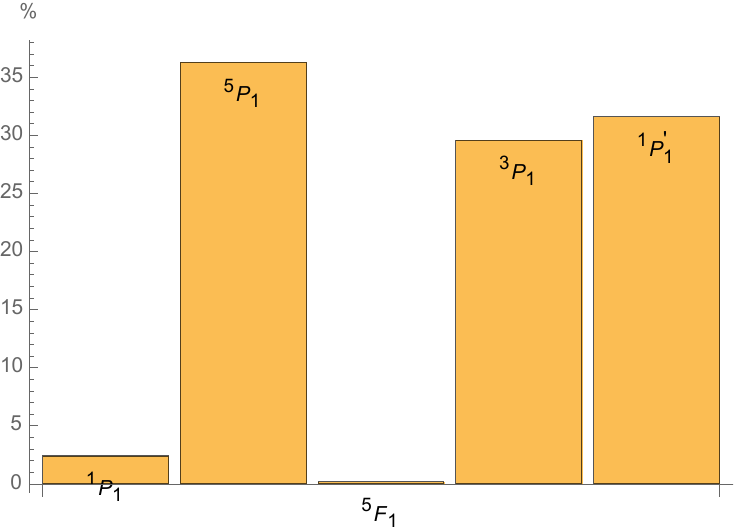}
   \caption{Typical $\Phi$ radial 
   wavefunctions in the mixed $J^{PC}=1^{--}$ channel (upper plot), 
   for the lowest bound state for the bottom exotic state with $C=-3/4$ (isosinglet), in units of $\Lambda=1$ GeV. 
   The corresponding percentages content of the bound wavefunction in the $J^{PC}=1^{--}$ channel are  shown below, with their spectroscopic assignments.}
   \label{fig_fbj1-1--}
  \end{center}
\end{figure}

 \begin{figure}[h!]
 \begin{center}
  \includegraphics[width=6cm]{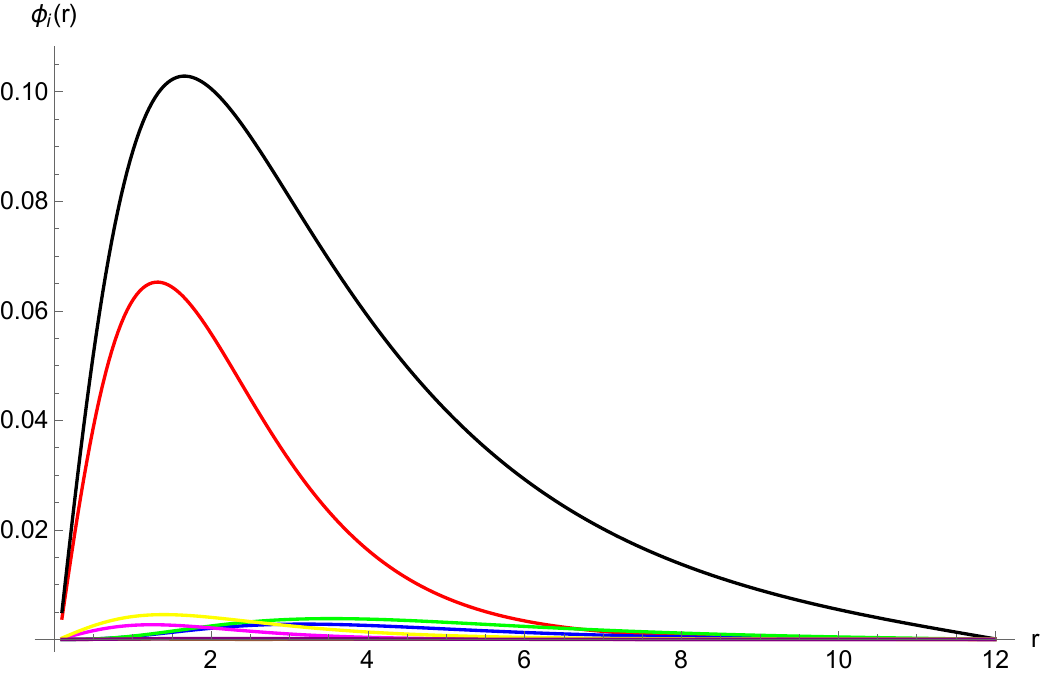}
   \includegraphics[width=6cm]{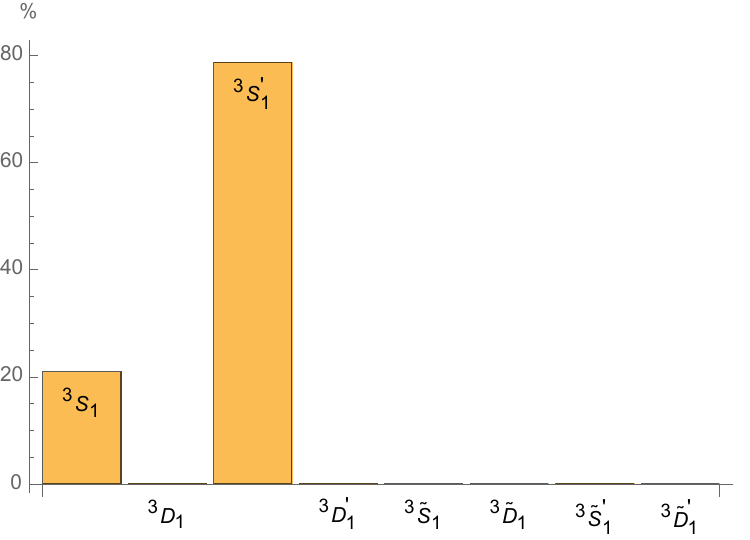}
   \caption{Typical $\Phi$ radial 
   wavefunctions in the mixed $J^{PC}=1^{+-}$ channel (upper plot), 
   for the lowest bound state for the bottom exotic state with $C=+1/4$ (isotriplet), in units of $\Lambda=1$ GeV. 
   The corresponding percentages content of the bound wavefunction in the $J^{PC}=1^{+-}$ channel are shown below, with their spectroscopic assignments.}
   \label{fig_fbj1-(1)1+-}
  \end{center}
\end{figure}

In Fig.~\ref{fig_fbj1-1--} we show the behavior of the typical isosinglet bound wavefunctions
contributing in the $J^{PC}=1^{--}$ channel with energy $E=10.558$ GeV ($\Lambda=1 $ GeV)
(upper-plots). The percentage content of the same wavefunction is displayed as a histogram with the appropriate 
parity labels in the lower display, with the ${}^SL_J$ assignments referring to the $(0^-,1^-)$ multiplet
only. From the assignments given in (\ref{J1LSJ}),
we see that the $1^{--}$ isosinglet bound state is mostly P-wave with  equal admixture of  $B^*\bar B^*$ (${}^5P_1$), 
$B\bar B^*$ (${}^3P_1$)  and $B\bar B$ (${}^1P_1^\prime$) molecules. We note the clear repulsion of
the P-waves near the origin.
The molecules are mostly from the $(0^-,1^-)$ multiplet as those from the $(0^+,1^+)$  decouple and unbind. 
The effect of the latters is to cause the formers to bind twice more. 
This isosinglet molecular exotic is well below the reported $Y_b(10888)$.

In Fig.~\ref{fig_fbj1-(1)1+-} we show the behavior of the typical isotriplet bound wavefunctions
contributing in the $J^{PC}=1^{+-}$ channel with energy $E=10.592$ GeV ($\Lambda=1 $ GeV)
(upper-plots). The percentage content of the same wavefunction is displayed as a histogram with the appropriate 
parity labels in the lower display, with the ${}^SL_J$ assignments referring to the $(0^-,1^-)$ multiplet, 
and the ${}^S\tilde L_J$ assignments referring to the $(0^+,1^+)$ multiplet. From the assignments given in (\ref{J1LSJ}),
we see that the $1^{+-}$ mixed isotriplet bound state is mostly an S-state 
made primarily of $B\bar B^*$ (${}^3S^\prime_1$) molecules  with a small admixture of 
$B^*\bar B^*$ (${}^3S^\prime_1$) molecules.
The molecules are mostly from the $(0^-,1^-)$ multiplet as those from the $(0^+,1^+)$ again decouple and unbind. 
We identify this exotic molecule as a mixed state of  the reported pair of isotriplet exotics 
$Z^+_b(10610)$ and $Z^+_b(10650)$.

\section{Conclusions}

We have analyzed molecular states of doubly heavy mesons mediated by one-pion exchange for both the
chiral parteners $(0^\pm , 1^\pm)$ as a coupled channel problem, for all the molecular configurations with
$J\leq 1$. Our  results
show that  the binding energy is sensitive to the cutoff used for the one-pion exchange interaction which is
substantial in the lowest partial waves. All other parameters are fixed by symmetry and data. 
Our results complement and extend those presented 
in~\cite{THORSSON,KARLINER,OTHERS,OTHERSX,OTHERSZ,OTHERSXX} by taking into 
account the strictures of chiral and heavy quark symmetry, and by retaining most coupled channels between the
$(0^-,1^-)$ multiplet and its chiral partner $(0^+,1^+)$. The key aspect of this coupling is to cause the
molecules in the $(0^-,1^-)$ multiplet to bind about twice more, and the molecules in the $(0^+,1^+)$ multiplet to unbind.

For channel couplings with $J\leq 1$, we have found that only the charm isosinglet exotic molecules with $J^{PC}=1^{++}$
is strictly bound for a pion-exchange cutoff $\Lambda=1 $ GeV. This state is identified with the reported isosinglet exotic 
X(3872) which in our case is mostly an isosinglet  $D\bar D^*$ molecule in the ${}^1S_0$ channel with no D-wave admixture. 
The attraction in the isotriplet channel with $J^{PC}=1^{+-}$ is too weak to bind the $D\bar D^*$ compound, suggesting that
the reported isotriplet $Z_C(3900)^\pm$ is at best a near treshold resonance.
All other $J^{PC}$ assignments with charm for both the isotriplet and isosinglet  are unbound. The noteworthy absence 
in our analysis of the $Y(4260)$, $Y(4360)$ and $Y(4660)$ may point to the possibility of their constituents made of excited
$(D_1,D_2)$ heavy mesons and their chiral partners~\cite{MACIEK,EXCITED}, which we have not considered. 

In contrast, and for the same choice of the cutoff,  we have identified
several isosinglet bottom exotic molecules in the $J^{PC}=0^{\pm +}, 1^{+\pm}, 1^{--}$ channels which are mostly admixtures of
the heavy-light mesons in the $(0^-,1^-)$ multiplet. We have only found one isotriplet bottom exotic molecule with $J^{PC}=1^{+-}$ 
which we have identified with the pair $Z^+_b(10610)$ and $Z^+_b(10650)$, which is a mixed state in our analysis.
The isosinglet bottom exotic molecule with $J^{PC}=1^{++}$ is a potential candidate for $X_b(10532)$, the bottom analogue
of the charm exotic $X(3872)$. 

Our results show that the cross chiral mixing between the $(0^\pm, 1^\pm)$ multiplets while strong, does not generate
new mixed molecules of the type $D{\bar {\tilde D}}^*$ and alike, as suggested in~\cite{OTHERS}. Rather,
it prevents the formation of dual chiral molecules of the type $\tilde D {\bar {\tilde D}}^*$ and alike, which would be
otherwise possible. In the process, it provides for a stronger binding of the low lying molecules in the $(0^-, 1^-)$
multiplet in comparison to the results in~\cite{OTHERSX}. Most noteworthy, is the appearance of a single
bound isosinglet $J^{PC}=1^{++}$ charm exotic molecule in our analysis, with also one single bound isotriplet bottom 
exotic molecule with $J^{PC}=1^{+-}$ but several isosinglet bottom exotic molecular states with 
$J^{PC}=0^{\pm +}, 1^{+ \pm}$. The latters may transmute to broad resonances by mixing to bottomia 
with similar quantum numbers.

Clearly higher values of $J>1$ may also be considered using the same construct, 
but the molecular configurations maybe too large to bind, a point in support of their 
absence in the currently reported experiments. The recoupling of the current bound
state problem to the open channels with charmonia and bottomia is also  important to
consider,  but requires a more extensive analysis of the multi-channel scattering problem.
Finally, the extension
of the present analysis to $D_s$ and $B_s$ molecules using the heavier eta-exchange~\cite{SKARLINER}, as well
as exotic baryonic molecules shoud be of interest in light of the ongoing  experimental programs.

\section{Acknowledgements}

This work was supported by the U.S. Department of Energy under Contract No.
DE-FG-88ER40388.

 \vfil

\end{document}